\documentclass[journal]{IEEEtran}
\usepackage{cite}
\usepackage{amsmath,amssymb,amsfonts}
\usepackage{algorithmic}
\usepackage{graphicx}
\usepackage{algorithm,algorithmic}
\usepackage{hyperref}
\hypersetup{hidelinks=true}

\usepackage{textcomp}
\usepackage{booktabs}
\usepackage{tabularx}
\usepackage{multirow}
\usepackage{adjustbox}
\begin{document}

\title{Physics-Grounded Objective-Conditioned Deep Beamforming for Adaptive Resolution-Contrast Control in Plane-Wave Ultrasound Imaging}

\author{Gopika Gopikrishnan,~\IEEEmembership{Student Member,~IEEE},
Mahesh Raveendranatha Panicker,~\IEEEmembership{Senior Member,~IEEE}
\thanks{Gopika Gopikrishnan and Mahesh Raveendranatha Panicker are with the Singapore Institute of Technology, Singapore (e-mail: 2403938@sit.singaporetech.edu.sg). This work was supported by the Ministry of Education, Singapore, and the SIT x NVIDIA AI Centre.}
\thanks{This is a preprint. It has not been peer-reviewed and published in a journal. The proposed dataset, source code, and trained weights will be made publicly available.}}
\IEEEaftertitletext{\vspace{-2em}}
\maketitle

\begin{abstract}
Plane-wave ultrasound enables ultrafast imaging but remains constrained by a receive-aperture-dependent resolution-contrast trade-off. Fine structures benefit from high spatial resolution, whereas diffuse tissue and lesions often require improved contrast and clutter suppression. These objectives favor different receive-aperture configurations, yet conventional beamformers operate at a fixed imaging point. We propose a physics-grounded, objective-conditioned deep beamformer that enables controllable image formation within a single network. From single-plane-wave channel data, the network predicts spatially varying receive-apodization weights, conditioned on the desired imaging objective through feature-wise linear modulation. The conditioning variable is explicitly tied to a physically defined receive-aperture configuration and supervised using the corresponding coherent-plane-wave-compounding target. Differentiable, region-of-interest-aware, physics-guided quality losses further promote objective-specific resolution and contrast during end-to-end training. The model is trained exclusively on simulated channel data spanning diverse target morphologies and echogenicities, and evaluated zero-shot on the PICMUS and CUBDL datasets, including unseen probe geometries. The proposed framework generally improves upon single-plane-wave delay-and-sum reconstruction while providing objective-dependent control. On PICMUS, the proposed method improves gCNR by up to $16.8\%$ and axial resolution by $22\%$ compared with the reference beamforming method. Moreover, switching between the contrast- and resolution-oriented objectives enables controllable changes of up to $4.9\%$ in lesion contrast and $11.4\%$ in lateral resolution. Compared with the best-performing learning-based baselines, the proposed
method reduces lateral full-width-at-half-maximum by $5.9\%$ and improves
generalized contrast-to-noise ratio by $6.6\%$.
\end{abstract}

\section{Introduction}

\IEEEPARstart{U}{ltrafast} plane-wave ultrasound has transformed diagnostic imaging by enabling frame rates of several thousand images per second, making real-time visualization of rapidly changing physiological processes possible~\cite{sandrin1999time,montaldo2009coherent,tanter2014ultrafast}. 
Coherent plane-wave compounding (CPWC) has been proposed to improve lateral resolution, suppress clutter, and enhance tissue contrast~\cite{montaldo2009coherent,couture2012ultrasound} with a reduction in achievable frame rate. Reconstructing CPWC-quality images from a single transmission has become one of the central challenges in modern ultrasound beamforming.
Receive aperture and apodization introduce a fundamental trade-off between spatial resolution and contrast~\cite{savoia2014improved, holmes2024f}. Large apertures narrow the point-spread function, improving visualization of small structures, whereas narrower apodization enhances contrast and reduces clutter at the cost of lateral resolution~\cite{trahey1986speckle,perrot2021so}. Thus, no single receive configuration is optimal across anatomies and tasks, yet conventional beamforming typically uses a fixed aperture or f-number ($F\#$) strategy. Recent deep learning (DL)-based beamformers have demonstrated remarkable improvements over conventional delay-and-sum beamforming (DAS-B) by learning adaptive image reconstruction directly from raw channel RF data~\cite{luchies2018deep}. Existing approaches can be broadly classified into three categories: 1) Image-to-image methods which enhance conventionally beamformed images through post-processing networks~\cite{gasse2017high}, 2) signal-to-image methods which directly reconstruct B-mode images from RF data using end-to-end neural networks and 3) adaptive beamforming methods which predict receive-apodization weights directly from channel data~\cite{luchies2018deep,luijten2020adaptive,hyun2019beamforming}. Most deep beamformers learn a fixed policy and therefore operate at a single resolution-contrast trade-off across anatomies. This conflicts with the clinical needs; vascular imaging may prioritize lumen contrast and clutter suppression, musculoskeletal imaging may prioritize higher resolution tissue boundaries and fascicular structures, and breast imaging of lesion may prioritize higher contrast. These tasks require objective-specific beamforming rather than a universal compromise. Recent studies make neural beamformers selectable rather than fixed; synthesizing DAS, minimum variance (MV), or despeckled reconstructions from one model~\cite{khan2021switchable}, or learning spatially adaptive weighting jointly with downstream objectives~\cite{li2026dp,amar2025deep}. These developments highlight the need for a physically grounded neural beamforming framework that explicitly addresses the resolution-contrast trade-off and provides controllable image formation according to the desired imaging objective.

In this work, we introduce a physics-grounded objective-conditioned adaptive beamformer that explicitly represents the desired imaging objective as a controllable beamforming condition. To the best of our knowledge, this is the first attempt to explicitly condition a deep learning-based beamformer towards counter objectives of higher resolution and contrast. The main contributions of this work are as follows.
\begin{enumerate}
\item A novel physics-grounded, objective-conditioned DL beamformer that represents the receive-aperture resolution-contrast trade-off as a controllable imaging condition.
\item A physically constrained, data-adaptive apodization mechanism that predicts receive weights constrained to the probability simplex.
\item A novel weakly supervised, differentiable, region-of-interest (ROI)-aware physics-guided training loss function.
\item A channel-domain processing pipeline tailored to the characteristics of ultrasound RF signals and spatial sampling, using polarity-preserving RF feature extraction, anisotropic feature processing, and patch-wise learning of spatially varying Time-of-Flight Corrected (ToFC) channel data.
\item Zero-shot cross-domain generalization demonstrated by training only on controlled simulated channel data with diverse tissue geometries and target echogenicities, followed by direct evaluation on independent PICMUS and CUBDL datasets, including previously unseen transducer geometries and without real-data fine-tuning.
\end{enumerate}
\begin{figure}
    \centering
    \includegraphics[width=\linewidth]{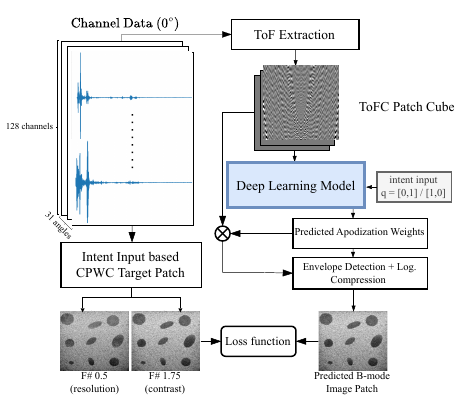}
    \caption{Overview of the objective-conditioned beamforming framework. Time of flight (ToF) extraction transforms a single unsteered ($0^\circ$), 128-channel RF acquisition into a ToFC patch cube, which is combined with an objective vector \(q\) (\([1,0]\): resolution; \([0,1]\): contrast). The network predicts pixel-wise receive-apodization weights, followed by envelope detection and log compression to form the B-mode patch. Training uses objective-specific, aperture-matched 31-angle CPWC targets, whereas inference requires only the \(0^\circ\) acquisition.}
    \label{fig:overview} 
    \end{figure}
    
\section{Methods}
An overview of the proposed framework is shown in Fig.~\ref{fig:overview}. Given an SPW acquisition, the received RF data are transformed into ToFC representations. A novel conditioning variable 
$q$ encodes the desired imaging objective and modulates a shared encoder-decoder through Feature-wise Linear Modulation (FiLM)~\cite{perez2018film}, selecting between operating regimes optimized for spatial resolution or lesion/lumen contrast. The predicted weights are applied through the conventional receive beamforming pipeline, and each objective is supervised during training with an aperture-matched 31-angle CPWC target under pixel-wise, perceptual, and physics-guided losses. It has to be noted that, we employ a patch-wise input driven by depth instead of the full ToFC data input which results in low complexity and depth-agnosticity as in our previous work \cite{gopikrishnan2026adapt}.

\subsection{Network Architecture}
\label{sec:arch}
\begin{figure*}[t]
    \centering
    \includegraphics[width=0.9\linewidth, trim = {0 0.5cm 0 0}, clip]{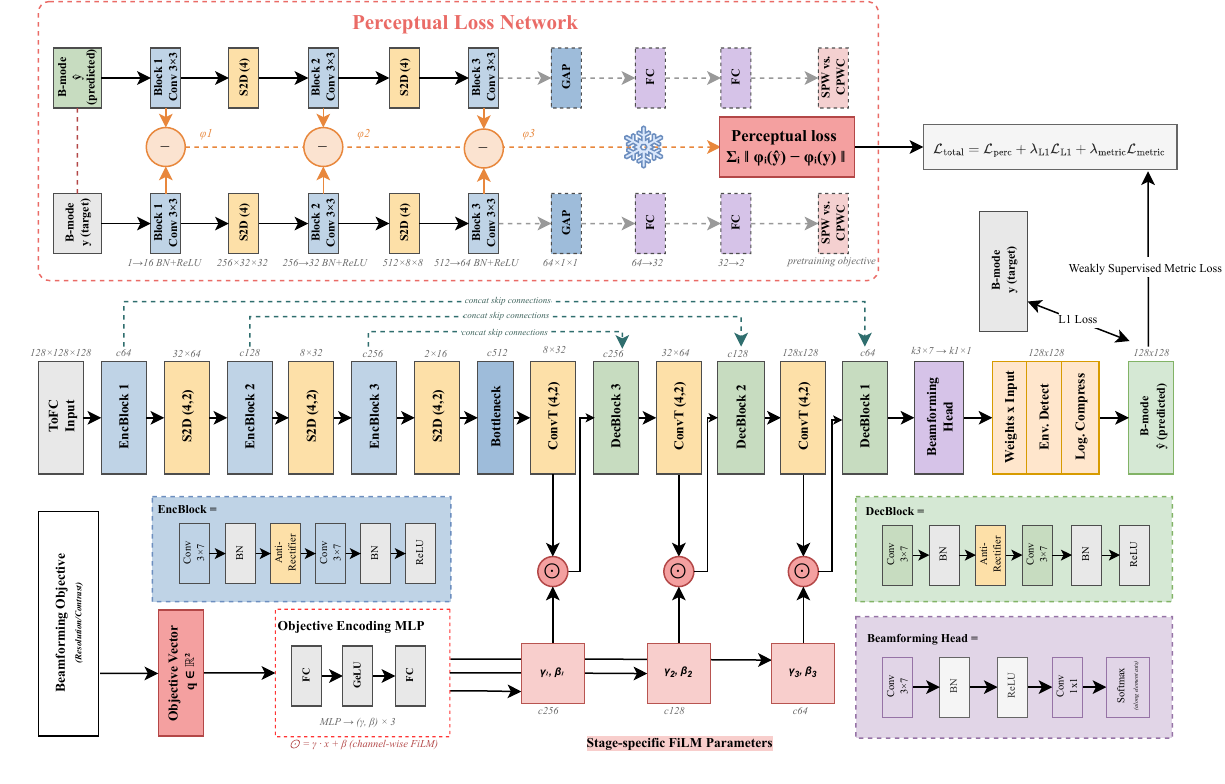}
    \caption{Architecture of the proposed objective-conditioned adaptive
beamformer ($c, k$ denote the number of channels and kernel size respectively). The ToFC patch cube passes through a
three-stage encoder with anisotropic
space-to-depth downsampling, a $512$-channel bottleneck, and a
symmetric decoder with transposed-convolution upsampling
and concatenated skip connections. Each encoder and decoder block applies
convolution-BN-AntiRectifier followed by convolution-BN-ReLU with $3\times7$
kernels. The imaging objective $q\in\mathbb{R}^2$ is mapped by an objective encoding MLP to stage-specific FiLM parameters $(\gamma_k,\beta_k)$ that modulate the three decoder stages channel-wise ($\odot:\gamma\!\cdot\!x+\beta$). The beamforming head outputs per-pixel receive weights, which are applied to
the input ToFC data before conventional beamforming and post-processing. The upper dashed branch is the training-only perceptual loss network: a frozen SPW-vs-CPWC classifier whose features $\varphi_1,\varphi_2,
\varphi_3$ give the perceptual term. The total loss combines perceptual, $L_1$,
and weakly-supervised metric losses.}
    \label{fig:model_architecture}
\end{figure*}
The proposed beamformer maps the ToFC tensor $\mathbf{S}\in\mathbb{R}^{N\times H\times W}$ to a spatially varying
receive-apodization field $\mathbf{W}\in\mathbb{R}^{N\times H\times W}$. The encoder extracts multiscale features using anisotropic S2D operations, while the decoder reconstructs the apodization field through transposed convolutions and skip connections. Objective conditioning is introduced through stage-specific FiLM modulation
in the decoder, and a final head produces normalized receive-element weights.

\subsubsection{Encoder}
The encoder comprises three feature-extraction blocks with output widths
$64$, $128$, and $256$. Each block consists of two
convolution-normalization stages, with the first convolution followed by an
antirectifier~\cite{shang2016understanding} and the second followed by ReLU. 
The output of each block is retained as a skip feature and is
then passed through anisotropic Space-to-Depth (S2D) downsampling before entering the next
encoder stage. Thus, the three S2D operations progressively reduce the
spatial resolution while increasing the feature-channel dimension.

\subsubsection{Anisotropy-Matched Downsampling and Kernels}
Each patch spans $3.24$~mm axially and the full $38.37$~mm aperture width on a $128 \times 128$ grid, yielding a spatial aspect ratio of approximately $12:1$. Consequently, the axial axis is heavily oversampled ($\approx 7\times$) relative to the near-critically sampled lateral axis. To handle this spatial anisotropy, we adopt two design choices.

First, downsampling is performed via lossless anisotropic S2D rearrangement \cite{shi2016real}:
\begin{equation}
\mathrm{S2D}:\ (C, H, W) \mapsto \big(C f_h f_w,\ \tfrac{H}{f_h},\ \tfrac{W}{f_w}\big)
\label{eq:s2d}
\end{equation}
with downsampling factors $(f_h, f_w) = (4, 2)$, followed by a learned pointwise convolution. Standard max-pooling retains only the peak sample per window, discarding axial timing and phase where adjacent fast-time samples lie within a wavelength. S2D reduces the loss of spatial information during downsampling by folding spatial dimensions into the channel dimension, while retaining local phase relationships relevant to coherent beamforming. Downsampling the axial axis more aggressively than the lateral axis ($f_h > f_w$) reduces axial redundancy while preserving lateral spatial resolution.

Second, we use $3 \times 7$ rectangular convolution kernels instead of standard square kernels. Because the lateral dimension undergoes less downsampling, the wider kernels provide a larger lateral receptive field for capturing echo coherence across the aperture, while the smaller axial extent helps retain axial detail when estimating the apodization weights.

\subsubsection{Bottleneck}
The three encoder stages are followed by a bottleneck at the coarsest spatial resolution, with 512 feature channels. It consists of two convolution-normalization-antirectifier stages followed by dropout ($p=0.2$). The bottleneck captures the highest-level features used by the decoder to reconstruct the spatially varying receive-apodization field.

\subsubsection{Imaging-Objective Encoding}
\label{sec:objective_encoding}

To guide the beamformer toward either higher contrast or higher spatial resolution, the desired imaging objective is provided to the network as a low-dimensional conditioning vector $q$. Instead of directly concatenating $q$ with the spatial feature maps, a shallow multilayer perceptron (MLP) transforms $q$ into a compact embedding that is used to modulate the network, as described in Section~\ref{sec:intent}.

\subsubsection{Decoder}
The decoder mirrors the three encoder stages with feature widths of $256$, $128$, and $64$. Each decoder stage first performs learnable anisotropic upsampling using a transposed convolution matched to the corresponding S2D factor, followed by concatenation with the feature map from the corresponding encoder stage through a skip connection. The concatenated features are then processed by the decoder convolutional block. At each stage, the feature representation is modulated by stage-specific FiLM parameters generated by the imaging-objective encoder. This provides three objective-dependent feature transformations along the decoder pathway. Conditioning is therefore applied after feature extraction and at progressively finer spatial resolutions, allowing the specified imaging objective to influence the predicted apodization field at multiple scales. The decoder outputs the receive-apodization weight tensor.
\subsubsection{Beamforming head}
\label{sec:head}

The final decoder feature map is passed to the beamforming head, which
consists of a $3\times7$ convolution-normalization-ReLU block followed by a
$1\times1$ convolution. The latter produces $N$ logits
$z_e(\mathbf{r})$ corresponding to the $N$ receive elements at every image
location. A softmax is applied along the receive-element dimension:
\begin{equation}
w_e(\mathbf{r}) = \frac{\exp\!\big(z_e(\mathbf{r})\big)}
{\sum_{e'=1}^{N}\exp\!\big(z_{e'}(\mathbf{r})\big)},
\qquad \sum_{e=1}^{N} w_e(\mathbf{r}) = 1,\; w_e(\mathbf{r}) \ge 0
\label{eq:softmax}
\end{equation}
Equation~\eqref{eq:softmax} constrains the predicted receive-apodization weights at each pixel to the probability simplex $\Delta^{N-1}$, ensuring non-negative, unit-sum weights. Consequently, the network learns a data- and imaging-objective-dependent apodization within this physically constrained space. The unit-sum constraint maintains a consistent overall receive gain while allowing the predicted weight distribution to adaptively shape the sidelobe structure. The non-negative, unit-sum weighting in~\eqref{eq:softmax} provides a stable, data-adaptive apodization that can suppress individual channels without relying on the signed-weight destructive interference used in MV beamforming.

\subsubsection{Perceptual loss branch}
The upper branch in Fig.~\ref{fig:model_architecture} denotes the
training-only perceptual loss network. The predicted and target B-mode images are passed through the same frozen feature extractor, and their intermediate feature representations are compared
to compute a domain-specific perceptual loss.
The perceptual term \cite{johnson2016perceptual} $\mathcal{L}_{\mathrm{perc}}$ in~\eqref{eq:total} is computed using a feature extractor trained to distinguish SPW from CPWC reconstructions. Because the extractor is trained on this distinction, its intermediate features are expected to capture image characteristics that differ between the two reconstruction modes, including resolution, clutter, and texture. This provides a perceptual representation aligned with the reconstruction differences that the proposed beamformer aims to reproduce. Given the frozen extractor $\phi$ with layers
$\ell \in \mathcal{P}$, the perceptual loss is the feature-space distance between
prediction and target,
\begin{equation}
\mathcal{L}_{\mathrm{perc}} = \sum_{\ell \in \mathcal{P}}
\lambda_\ell \big\| \hat{\phi}_\ell(\hat{B}) - \hat{\phi}_\ell(B) \big\|_1 ,
\label{eq:perc}
\end{equation}
where $\hat{\phi}_\ell$ denotes the $L_2$-channel-normalized features at tap
$\ell$ and $\lambda_\ell$ are per-tap weights. This branch is used only during training and is not included in the inference-time beamforming pathway.
\subsection{Imaging Objectives and Aperture Selection}
\label{sec:intent}
\subsubsection{Receive Aperture based Conditioning}

The receive f-number, $F\# = z/A$, determines the active aperture width $A$ at depth $z$ and influences the resolution-contrast trade-off. A larger aperture (lower $F\#$) generally provides higher lateral resolution, whereas a smaller aperture (higher $F\#$) can reduce off-axis contributions and improve contrast~\cite{perrot2021so}. An $F\#$ sweep was performed and $F\#=0.5$ was selected as the resolution-oriented operating point based on the finest pin-target resolution, and $F\#=1.75$ as the contrast-oriented operating point based on the contrast-target results. These two empirically selected points define the endpoints of the conditioning variable.

\begin{equation}
q=
\begin{cases}
[1,0], & \text{resolution objective } (F\#=0.5)\\
[0,1], & \text{contrast objective } (F\#=1.75)
\end{cases}
\label{eq:objectives}
\end{equation}
with training targets reconstructed using dynamic receive apertures $A(z)=z/F\#$ for the corresponding fixed $F\#$. At inference, the imaging
objective is continuously parameterized as
\begin{equation}
q(t)=[1-t,t], \qquad t\in[0,1]
\label{eq:objective_interpolation}
\end{equation}
where $t=0$ corresponds to the resolution objective and $t=1$ to the contrast objective. Intermediate values of $t$ are not independently supervised and generate a smooth family of receive-apodization fields between the two supervised operating points; their behavior is discussed in Section~\ref{sec:discussion}.

\subsubsection{Imaging-objective conditioning}
The imaging objective is injected through FiLM \cite{perez2018film}, allowing a single beamformer to adapt its apodization to different imaging objectives without requiring separate network parameters for each objective. Given decoder feature map $\mathbf{h}_l \in \mathbb{R}^{C_l \times H_l \times W_l}$ at level $l$, FiLM applies a feature-wise affine transformation,
\begin{equation}
\mathbf{h}_l \leftarrow
\boldsymbol{\gamma}_l(q)\odot\mathbf{h}_l
+\boldsymbol{\beta}_l(q)
\label{eq:film}
\end{equation}
where $q$ specifies the desired imaging objective and the modulation parameters
$\boldsymbol{\gamma}_l,\boldsymbol{\beta}_l\in\mathbb{R}^{C_l}$ are generated by lightweight two-layer MLPs. To preserve the behavior of the unconditioned network at initialization, the FiLM scale is parameterized in residual form,
$\boldsymbol{\gamma}_l(q)=\mathbf{1}+\Delta\boldsymbol{\gamma}_l(q)$, and the final MLP layers are zero-initialized. At the start of training, this gives $\boldsymbol{\gamma}_l=\mathbf{1}$ and $\boldsymbol{\beta}_l=\mathbf{0}$, allowing objective-dependent modulation to be learned progressively while maintaining stable optimization. The encoder therefore extracts representations from the measured ToFC data before the imaging objective is introduced, while the decoder transforms these representations into objective-specific receive-apodization weights. FiLM is applied only in the decoder, restricting the effect of the imaging objective to the pathway responsible for aperture synthesis.

\subsubsection{Objective dropout and the neutral operating point}
To encourage smooth behavior between the supervised operating points, we adopt
classifier-free conditioning dropout~\cite{ho2022classifier}. During training,
the conditioning vector is replaced with the neutral objective
$q=[0.5,0.5]$ with probability $p$,

\begin{equation}
q_{\mathrm{eff}}
=
m\odot q
+
(1-m)[0.5,0.5]
\label{eq:cfg}
\end{equation}

where $m\sim\mathrm{Bernoulli}(1-p)$ is sampled independently for each training example. Rather than using a zero conditioning vector, which does not correspond to any defined imaging objective, we use the midpoint $q=[0.5,0.5]$. Under the continuous parameterization $q(t)=[1-t,t]$, this corresponds to $t=0.5$ and represents an equal interpolation between the two supervised endpoints. Thus, the midpoint provides a defined intermediate objective rather than an artificial zero-conditioning state.

\subsection{Physics-Guided Differentiable Losses}
\label{sec:losses}
In addition to pixel-wise and perceptual losses for global image fidelity, we
introduce differentiable ROI-aware surrogates for lateral point-target
resolution and lesion/lumen contrast. The resolution surrogate is applied only
to the resolution-conditioned branch, while the contrast surrogate is applied
only to the contrast-conditioned branch. Let
$\hat{B}\in[0,1]^{H\times W}$ denote the predicted B-mode image after envelope
detection, log compression, and normalization. ROI locations, centers, and
radii are obtained from phantom annotations and mapped to image coordinates.

\subsubsection{Soft resolution loss}
The lateral full-width-at-half-maximum (FWHM) of a point target is
non-differentiable because it requires peak localization and threshold crossings. To obtain a differentiable surrogate, we approximate the mainlobe spread using its intensity-weighted second moment. For a Gaussian profile, FWHM and standard deviation are related by $\mathrm{FWHM}=2\sqrt{2\ln2}\,\sigma\approx2.355\sigma$, so minimizing the estimated spread provides a differentiable proxy for lateral FWHM. For a point
target whose intensity peak lies at $(x_c, z_c)$, a lateral profile is obtained
by averaging the three image rows nearest the peak along the axial ($z$) axis,
\begin{equation}
p(x)=\frac13\sum_{\delta=-1}^{1}\hat B(x, z_c+\delta),
\label{eq:profile}
\end{equation}

which suppresses sub-sample peak-localization noise while remaining within the
axial mainlobe, and is converted into peak-emphasizing weights

\begin{equation}
w(x)=\mathrm{ReLU}(p(x)-\bar p)+\epsilon
\label{eq:peakweight}
\end{equation}

where $\bar p$ is the profile mean. The weighted first and second moments are

\begin{equation}
\mu=\langle x\rangle_w,\qquad
\sigma^2=\langle(x-\mu)^2\rangle_w
\end{equation}

\begin{equation}
\mathcal L_{\mathrm{res}}
=
\frac{\sqrt{\sigma^2+\epsilon}}{W_p}
\label{eq:lres}
\end{equation}

where $\langle f(x)\rangle_w =
\frac{\sum_x w(x)f(x)}{\sum_x w(x)}$ denotes the normalized
intensity-weighted expectation. The peak-centered weighting emphasizes the mainlobe and a small positive contribution from the surrounding profile is retained through $\epsilon$.
The resulting second moment captures mainlobe spread and residual sidelobe
energy, providing a differentiable beamforming-oriented surrogate for FWHM. Minimizing $\mathcal L_{\mathrm{res}}$ reduces the lateral spread of point targets while remaining differentiable for end-to-end optimization.

\subsubsection{Soft contrast loss}
The lesion contrast is optimized using a differentiable surrogate for generalized
contrast-to-noise ratio (gCNR), defined as $\mathrm{gCNR}=1-\mathrm{OVLP}$, where $\mathrm{OVLP}$ denotes the overlap between the lesion and background intensity distributions. Because hard ROI masks and histogram binning are non-differentiable, both are replaced with smooth approximations.
Given the normalized elliptical distance
\begin{equation}
d(x,z)
=
\sqrt{
\left(\frac{x-x_0}{r_x}\right)^2
+
\left(\frac{z-z_0}{r_z}\right)^2}
\label{eq:elldist}
\end{equation}
 
the soft lesion and background masks are defined as,
\begin{equation}
m_{\mathrm{in}}
=
\sigma\!\left(\frac{0.8-d}{\tau}\right),
m_{\mathrm{out}}
=
\sigma\!\left(\frac{d-1.2}{\tau}\right)
\sigma\!\left(\frac{1.8-d}{\tau}\right)
\label{eq:masks}
\end{equation}
corresponding to the lesion interior and a surrounding annulus. $(x_0, z_0)$ is the center and $(r_x, r_z)$ are the lateral and axial
radii of the lesion ROI, such that $d=1$ on
the nominal lesion boundary.  The steepness parameter $\tau$ controls the sharpness of the sigmoid mask transitions. For a mask $m_i\in\{m_{\mathrm{in},i},m_{\mathrm{out},i}\}$, the corresponding mask-weighted kernel-density histogram is
\begin{equation}
h[k]
=
\sum_i
m_i
\exp\!\left(
-\frac{(v_i-c_k)^2}{2s^2}
\right),
h
\leftarrow
\frac{h}{\sum_k h[k]}
\label{eq:softhist}
\end{equation}
where $c_k$ denotes the center of histogram bin $k$ and $s$ controls the
kernel bandwidth. This kernel-density construction replaces hard histogram
assignment with a smooth function of the image intensities from which lesion and background distributions, $h_{\mathrm{in}}$ and $h_{\mathrm{out}}$, are obtained. The contrast loss is

\begin{equation}
\mathcal L_{\mathrm{con}}
=
\sum_{k=1}^{K}
\min
\left(
h_{\mathrm{in}}[k],
h_{\mathrm{out}}[k]
\right)
=
1-\widehat{\mathrm{gCNR}}
\label{eq:lcon}
\end{equation}
Minimizing $\mathcal L_{\mathrm{con}}$ reduces the overlap between lesion and
background intensity distributions while providing a piecewise-differentiable
surrogate through the soft masks and kernel-density estimation.

\subsubsection{Combined objective}
For the proposed approach, the physics-guided losses are applied only to the corresponding imaging objective, i.e., the $\mathcal L_{\mathrm{res}}$ supervises the resolution-conditioned branch using point-target ROIs, whereas $\mathcal L_{\mathrm{con}}$ supervises the contrast-conditioned branch using lesion ROIs. The overall training objective is

\begin{equation}
\mathcal L
=
\mathcal L_{\mathrm{perc}}
+
\lambda_1
\mathcal L_1(\hat B,B)
+
\lambda_m
\mathcal L_{\mathrm{metric}}
\label{eq:total}
\end{equation}

For the two supervised endpoint objectives, the corresponding physics-guided
surrogate is selected according to the conditioning state:

\begin{equation}
\mathcal L_{\mathrm{metric}}
=
\begin{cases}
\mathcal L_{\mathrm{res}}, & q=[1,0],\\
\mathcal L_{\mathrm{con}}, & q=[0,1]
\end{cases}
\label{eq:metricsel}
\end{equation}
Unlike pixel-wise and perceptual supervision alone, the proposed
objective-specific surrogates provide direct training signals for the
beamforming properties associated with each operating regime. 

\subsection{Training and Datasets}
\label{sec:training}

\subsubsection{Simulation dataset}
Training data are generated entirely in simulation using the SIMUS solver of the
MUST toolbox~\cite{garcia2021make}, configured for a linear array with $N = 128$
elements, centre frequency $f_c = 7.6$~MHz, and sampling frequency
$f_s = 30.4$~MHz. Seven target classes are simulated: point scatterers, bone interfaces,
anechoic cysts, hyperechoic cysts, hypoechoic cysts, linear vessels, and
round vessels. A multi-angle planewave transmit with 31 steered transmissions from $-18^\circ$ to $+18^\circ$ are then simulated to produce the raw RF channel data. From each simulated acquisition, the SPW ($0^\circ$) acquisition provides the network input and the coherently compounded acquisition provides the ground truth for training. Each acquisition is reconstructed at $F\#=0.5$ and $F\#=1.75$ using
\emph{ultraspy}~\cite{perrot2021so}, yielding aperture-matched target pairs. Both targets are generated for every scene regardless of its tissue class, so that the objective variable is statistically independent of the tissue content and the network must learn the objective as a control rather than as a proxy for tissue type. Fifty acquisitions are generated per class across the seven classes, yielding $350$ acquisitions, each divided into $9$ depth patches for a total of $3{,}150$ patches. Acquisitions are split $8{:}2$ at the acquisition level, yielding $280$ training and $70$ validation acquisitions, corresponding to $2{,}520$ and $630$ patches, respectively. All splits are performed at the acquisition level, so that patches from a single scatterer realization cannot appear in more than one split.

\subsubsection{Patch-wise training}
The network operates on ToFC channel patches of fixed size $128\times N_e$,
where 128 corresponds to the axial dimension (approximately $3.24$~mm) and
$N_e$ is the number of receive elements. Patch-wise processing enables
depth-agnostic learning by allowing the same network to operate on local
axial regions sampled from different depths, while substantially reducing
memory footprint and increasing the effective number of training samples as discussed in our previous work \cite{gopikrishnan2026adapt}.
During inference, adjacent axial patches are extracted with a 50\% overlap.
For each patch, the predicted receive-apodization weights are applied to the
corresponding ToFC samples through the beamforming operation, producing a beamformed RF patch. The overlapping beamformed RF patches are then assembled in their original spatial order to form the complete beamformed RF image. Envelope detection and log compression
are performed only after patch assembly, rather than independently on each
patch, thereby avoiding discontinuities and patch-boundary artifacts in the
final B-mode image.

\subsubsection{Optimization}
The network is trained using the AdamW optimizer with an initial learning rate
of $10^{-3}$ and a batch size of 4. The learning rate is adaptively reduced
using an \emph{ReduceLROnPlateau} scheduler over 200 training epochs. The learning rate is reduced by a factor of 0.5 when the validation loss does not improve for 10 consecutive epochs. Objective dropout is applied with probability $p=0.1$. Training is performed on a single NVIDIA RTX 4070 GPU. The resulting model contains 24.6M trainable parameters.

\subsubsection{Evaluation data}
The trained model is evaluated exclusively on unseen public datasets such as the PICMUS and CUBDL datasets~\cite{liebgott2016plane,bell2020challenge},
covering \emph{in-silico}, \emph{in-vitro}, and \emph{in-vivo} acquisitions, including carotid, breast, and musculoskeletal targets. The evaluation data span four linear-array probe geometries: L11-5v, L11-4v, L7, and L7-4, whose pitch, center frequency, and sampling rate differ from those of the simulated training array. Table~\ref{tab:probe_results} evaluates zero-shot cross-probe and cross-domain
generalization on previously unseen experimental and in-vivo acquisitions
without fine-tuning.

\section{Results}
\label{sec:results}
\begin{figure}
    \centering
    \includegraphics[width=\linewidth]{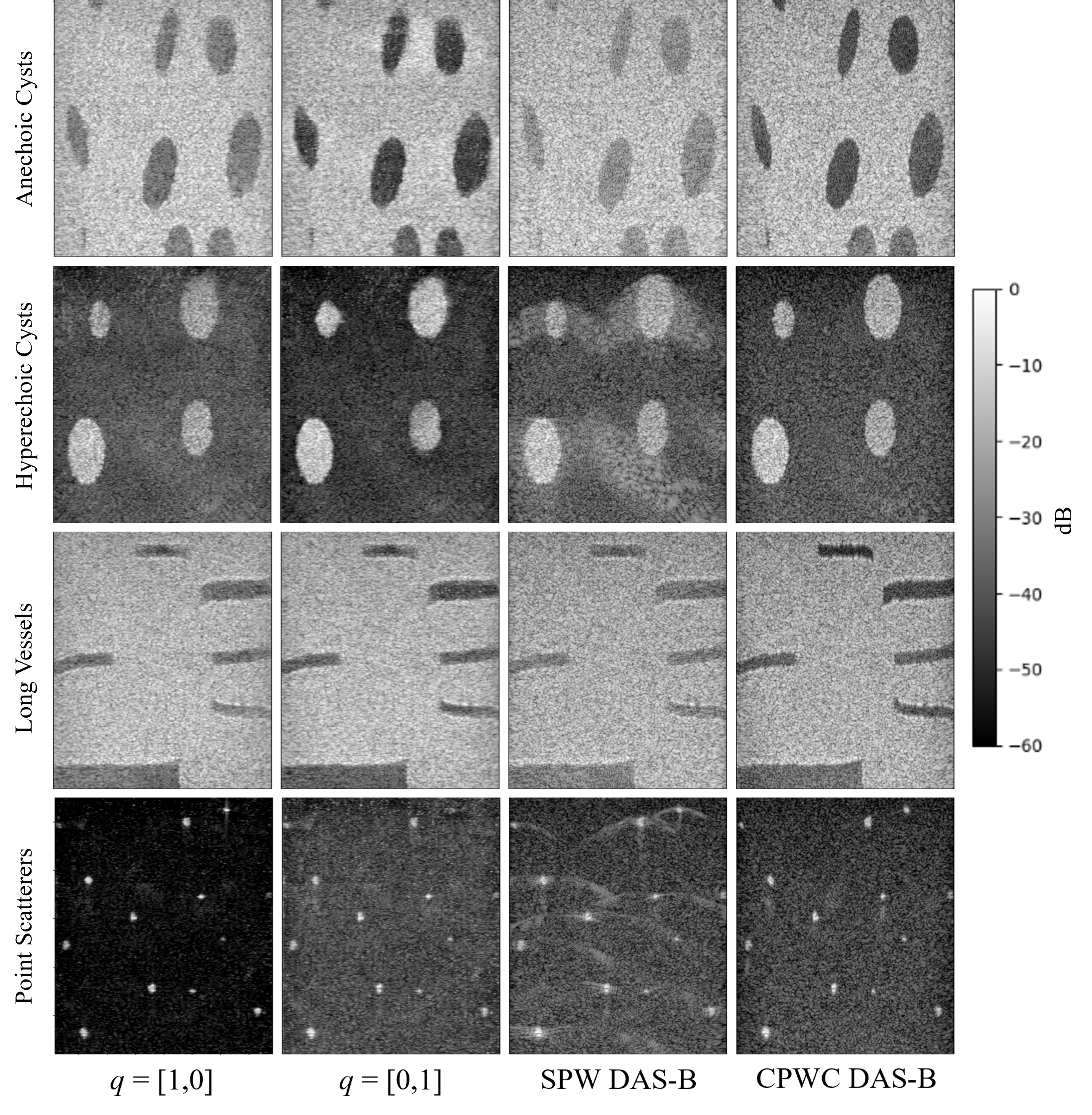}
    \caption{Qualitative results on simulated phantoms for the two imaging objectives, compared against the SPW and the CPWC DAS-B Images. Each row represents various tissue categories. The columns show the proposed
beamformer under the resolution objective $q=[1,0]$ and the contrast objective $q=[0,1]$, both reconstructed from the same SPW input. The contrast objective suppresses background clutter and improves lesion conspicuity in the
cyst targets, while the resolution objective produces sharper, more compact point
targets in the bottom row; both objectives markedly reduce the clutter and sidelobe artifacts visible in the SPW input, approaching the CPWC reference
while requiring only a single transmission.}
\label{fig:heldout_results}
\end{figure}
\begin{table}[hbtp!]
\caption{Performance on simulated targets at both imaging intents.}
\label{tab:categories}
\centering
\scriptsize
\setlength{\tabcolsep}{2.5pt}
\renewcommand{\arraystretch}{1.10}

\textbf{(a) Contrast performance}\par\smallskip
\begin{adjustbox}{max width=\columnwidth}
\begin{tabular}{@{}llcc@{}}
\toprule
Category & Method & gCNR$\uparrow$ & CNR (dB)$\uparrow$ \\
\midrule
\multirow{4}{*}{Anechoic}
& $q = [1,0]$ & $0.827\pm0.040$ & $1.89\pm0.14$ \\
& $q = [0,1]$ & $\mathbf{0.887\pm0.045}$ & $\mathbf{2.06\pm0.17}$ \\
& SPW     & $0.660\pm0.024$ & $1.56\pm0.05$ \\
& CPWC    & $0.918\pm0.019$ & $1.98\pm0.08$ \\
\midrule
\multirow{4}{*}{Hyperechoic}
& $q = [1,0]$ & $0.935\pm0.023$ & $2.16\pm0.10$ \\
& $q = [0,1]$ & $\mathbf{0.961\pm0.023}$ & $\mathbf{2.50\pm0.12}$ \\
& SPW     & $0.842\pm0.015$ & $2.12\pm0.03$ \\
& CPWC    & $0.943\pm0.013$ & $2.23\pm0.05$ \\
\midrule
\multirow{4}{*}{Hypoechoic}
& $q = [1,0]$ & $0.840\pm0.037$ & $1.93\pm0.11$ \\
& $q = [0,1]$ & $\mathbf{0.914\pm0.026}$ & $\mathbf{2.16\pm0.15}$ \\
& SPW     & $0.643\pm0.024$ & $1.53\pm0.07$ \\
& CPWC    & $0.887\pm0.025$ & $1.93\pm0.08$ \\
\midrule
\multirow{4}{*}{Cross vessels}
& $q = [1,0]$ & $0.741\pm0.097$ & $1.76\pm0.30$ \\
& $q = [0,1]$ & $\mathbf{0.798\pm0.105}$ & $\mathbf{1.96\pm0.31}$ \\
& SPW     & $0.566\pm0.058$ & $1.33\pm0.18$ \\
& CPWC    & $0.829\pm0.073$ & $1.85\pm0.22$ \\
\bottomrule
\end{tabular}
\end{adjustbox}

\smallskip
\textbf{(b) Resolution Performance}\par\smallskip
\begin{adjustbox}{max width=\columnwidth}
\begin{tabular}{@{}lcc@{}}
\toprule
Method & FWHM$_{\mathrm{lat}}$(mm)$\downarrow$& FWHM$_{\mathrm{ax}}$(mm)$\downarrow$ \\
\midrule
$q = [1,0]$ & $\mathbf{0.518\pm0.042}$ & $\mathbf{0.183\pm0.012}$ \\
$q = [0,1]$ & $0.568\pm0.057$ & $0.193\pm0.026$ \\
SPW     & $0.565\pm0.043$ & $0.210\pm0.014$ \\
CPWC    & $0.421\pm0.039$ & $0.208\pm0.017$ \\
\bottomrule
\end{tabular}
\end{adjustbox}
\end{table}

\begin{table}[hbtp!]
\centering
\caption{Performance comparison on PICMUS contrast and resolution phantoms.}
\label{tab:picmus}

\resizebox{\linewidth}{!}{%
\begin{tabular}{llcc}
\toprule
Phantom & Method & CNR (dB) $\uparrow$ & gCNR$\uparrow$ \\
\midrule
\multirow{4}{*}{\emph{In-silico}}
  & Proposed $q{=}[1,0]$ & $5.95\pm0.87$ & $0.947\pm0.035$ \\
  & Proposed $q{=}[0,1]$ & $\mathbf{6.47\pm1.38}$ & $\mathbf{0.980\pm0.012}$ \\
  & SPW   & $5.76\pm0.39$ & $0.839\pm0.026$ \\
  & CPWC        & $6.25\pm0.57$ & $0.977\pm0.012$ \\
\midrule
\multirow{4}{*}{\emph{In-vitro}}
  & Proposed $q{=}[1,0]$ & $5.13\pm0.31$ & $0.753\pm0.020$ \\
  & Proposed $q{=}[0,1]$ & $\mathbf{5.78\pm0.48}$ & $\mathbf{0.790\pm0.045}$ \\
  & SPW          & $4.90\pm0.22$ & $0.732\pm0.042$ \\
  & CPWC         & $6.87\pm0.40$ & $0.908\pm0.019$ \\
\bottomrule
\end{tabular}%
}

\resizebox{\linewidth}{!}{%
\begin{tabular}{llcc}
\toprule
Phantom & Method & FWHM$_{\text{lat}}$ (mm)$\downarrow$ & FWHM$_{\text{ax}}$ (mm)$\downarrow$ \\
\midrule
\multirow{4}{*}{\emph{In-silico}}
  & Proposed $q{=}[1,0]$ & $\mathbf{0.506\pm0.136}$ & $\mathbf{0.322\pm0.060}$ \\
  & Proposed $q{=}[0,1]$ & $0.519\pm0.136$ & $0.358\pm0.075$ \\
  & SPW          & $0.793\pm0.123$ & $0.413\pm0.007$ \\
  & CPWC        & $0.353\pm0.069$ & $0.491\pm0.056$ \\
\midrule
\multirow{4}{*}{\emph{In-vitro}}
  & Proposed $q{=}[1,0]$ & $\mathbf{0.815\pm0.118}$ & $\mathbf{0.488\pm0.054}$ \\
  & Proposed $q{=}[0,1]$ & $0.920\pm0.146$ & $0.498\pm0.076$ \\
  & SPW          & $1.151\pm0.264$ & $0.673\pm0.102$ \\
  & CPWC        & $0.768\pm0.082$ & $0.725\pm0.138$ \\
\bottomrule
\end{tabular}%
}
\end{table}

\begin{table}[hbtp!]
\caption{Out-of-distribution performance across transducers.}
\label{tab:probe_results}
\centering
\scriptsize
\setlength{\tabcolsep}{1.5pt}
\renewcommand{\arraystretch}{0.95}
\begin{tabular}{@{}llcccc@{}}
\toprule
Probe & Method & Lat. FWHM$\downarrow$ & Ax. FWHM$\downarrow$ & gCNR$\uparrow$ & CNR$\uparrow$ \\
& & (mm) & (mm) & & (dB) \\
\midrule
\multirow{4}{*}{L11-5v}
& $q=[1,0]$ & $\mathbf{0.879\pm0.174}$ & $\mathbf{0.445\pm0.060}$ & $0.758\pm0.039$ & $4.60\pm0.62$\\
& $q=[0,1]$ & $0.962\pm0.169$ & $0.511\pm0.100$ & $\mathbf{0.785\pm0.004}$ & $\mathbf{5.19\pm0.03}$\\
& SPW & $1.151\pm0.264$ & $0.673\pm0.102$ & $0.732\pm0.042$ & $4.90\pm0.22$\\
& CPWC & $0.768\pm0.082$ & $0.725\pm0.138$ & $0.908\pm0.019$ & $6.87\pm0.40$\\
\midrule
\multirow{4}{*}{L7}
& $q=[1,0]$ & $\mathbf{0.554\pm0.236}$ & $\mathbf{0.392\pm0.165}$ & -- & --\\
& $q=[0,1]$ & $0.597\pm0.275$ & $0.409\pm0.179$ & -- & --\\
& SPW & $0.388\pm0.070$ & $0.448\pm0.149$ & -- & --\\
& CPWC & $0.377\pm0.074$ & $0.450\pm0.150$ & -- & --\\
\midrule
\multirow{4}{*}{L7-4}
& $q=[1,0]$ & $\mathbf{0.585\pm0.143}$ & $\mathbf{0.573\pm0.147}$ & -- & --\\
& $q=[0,1]$ & $0.650\pm0.266$ & $0.611\pm0.147$ & -- & --\\
& SPW & $0.556\pm0.111$ & $0.617\pm0.143$ & -- & --\\
& CPWC & $0.478\pm0.153$ & $0.597\pm0.124$ & -- & --\\
\midrule
\multirow{4}{*}{L11-4v}
& $q=[1,0]$ & -- & -- & $0.432\pm0.180$ & $0.98\pm0.37$\\
& $q=[0,1]$ & -- & -- & $\mathbf{0.447\pm0.152}$ & $\mathbf{1.04\pm0.25}$\\
& SPW & -- & -- & $0.485\pm0.049$ & $1.00\pm0.07$\\
& CPWC & -- & -- & $0.464\pm0.159$ & $1.12\pm0.58$\\
\bottomrule
\end{tabular}
\end{table}

Table~\ref{tab:categories} reports the mean~$\pm$~s.d.\ over each category for the best checkpoint, with reconstructions in Fig.~\ref{fig:heldout_results}. The results show that across every contrast category, the contrast objective $q=[0,1]$ improves gCNR and CNR over the resolution objective $q=[1,0]$. The gain is largest on the hypoechoic cysts for gCNR, which rises by $8.8\%$, and on the hyperechoic cysts for CNR, which rises by $15.7\%$, whereas the remaining contrast categories improve by $2.8$-$7.7\%$ in gCNR and $9$-$11\%$ in CNR. The simultaneous gCNR and CNR gains confirm that these shifts reflect genuine lesion-background separation rather than a change in mean intensity. Conversely, on point targets, the resolution objective is the sharper of the two, narrowing the lateral FWHM by $8.8\%$ and the axial FWHM by $5.2\%$.

\subsection{Comparison with State of the Art}
\label{sec:sota}

\begin{figure*}[hbtp!]
    \centering
    \includegraphics[width=0.65\linewidth]{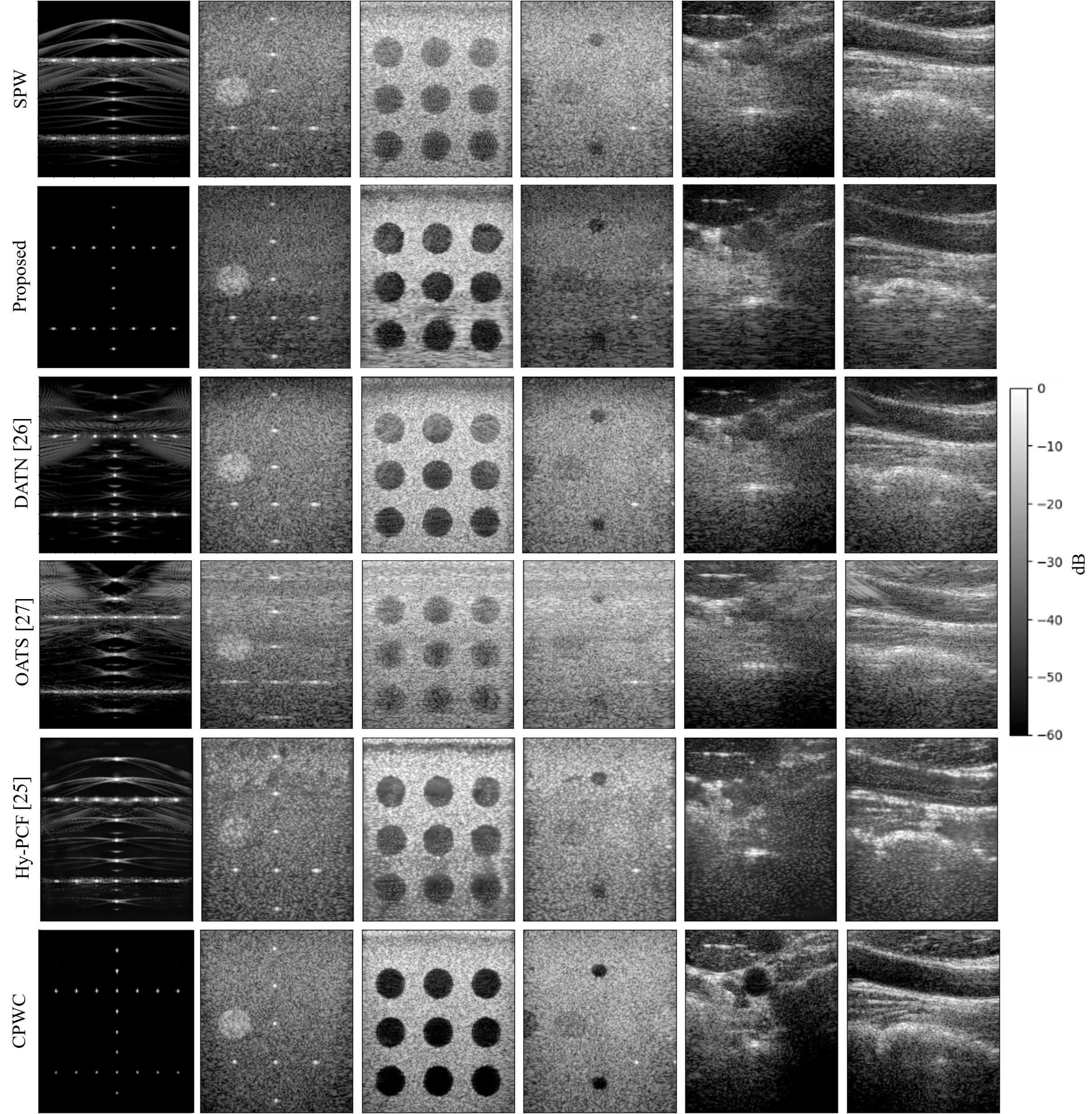}
    \caption{Qualitative comparison of beamforming methods on the PICMUS benchmark.
    Results are shown for the \emph{in silico}, \emph{in vitro}, and \emph{in vivo}
    datasets. Columns compare the SPW, CPWC, OATS~\cite{xiao2025optimized}, DATN~\cite{liu2026depth}, Hy-PCF~\cite{yu2026plane}, and the proposed method
    reconstructed under the resolution objective ($q=[1,0]$) and contrast
    objective ($q=[0,1]$). The proposed conditioned beamformer provides
    objective-dependent reconstruction from the same SPW acquisition,
    producing sharper point targets and anatomical boundaries under the resolution
    objective, while enhancing lesion conspicuity and suppressing background
    clutter under the contrast objective.}
    \label{fig:comparison}
\end{figure*}

Table~\ref{tab:sota} compares the proposed FiLM-conditioned adaptive beamformer with a few DL-based ultrasound reconstruction methods in the literature, including Hy-PCF~\cite{yu2026plane}, DATN~\cite{liu2026depth},
OATS~\cite{xiao2025optimized}. For the resolution objective, the proposed method attains the sharpest lateral
and axial responses among the compared learning-based beamformers, improving lateral FWHM by $5.9\%$ over the best baseline \cite{liu2026depth} and axial FWHM by $19.7\%$ over \cite{xiao2025optimized}. For the contrast objective, it achieves the highest gCNR, exceeding the best competing method \cite{yu2026plane} by $6.6\%$, while its CNR is within $0.1$~dB of \cite{liu2026depth}. The proposed method thus leads on resolution and lesion detectability from a single conditioned network, while maintaining agreement with the aperture-matched CPWC targets. Unlike DATN~\cite{liu2026depth}, which maps channel data directly to images, the proposed method predicts spatially varying receive-apodization weights from delay-aligned ToFC data and forms the image using conventional beamforming. OATS~\cite{xiao2025optimized} also learns apodization through differentiable beamforming, but targets a fixed objective. In contrast, our FiLM-conditioned shared model reconstructs the same SPW acquisition under different aperture regimes, enabling inference-time control of the resolution-contrast trade-off without retraining or separate networks. The qualitative results in Fig.~\ref{fig:comparison} show that the proposed method produces sharper point targets in the resolution mode while improving lesion contrast and suppressed sidelobes in the contrast mode, demonstrating that controllable adaptive beamforming provides greater flexibility than fixed reconstruction or fixed-apodization approaches.
\begin{figure*}[hbtp!]
    \centering
    \includegraphics[width=\linewidth]{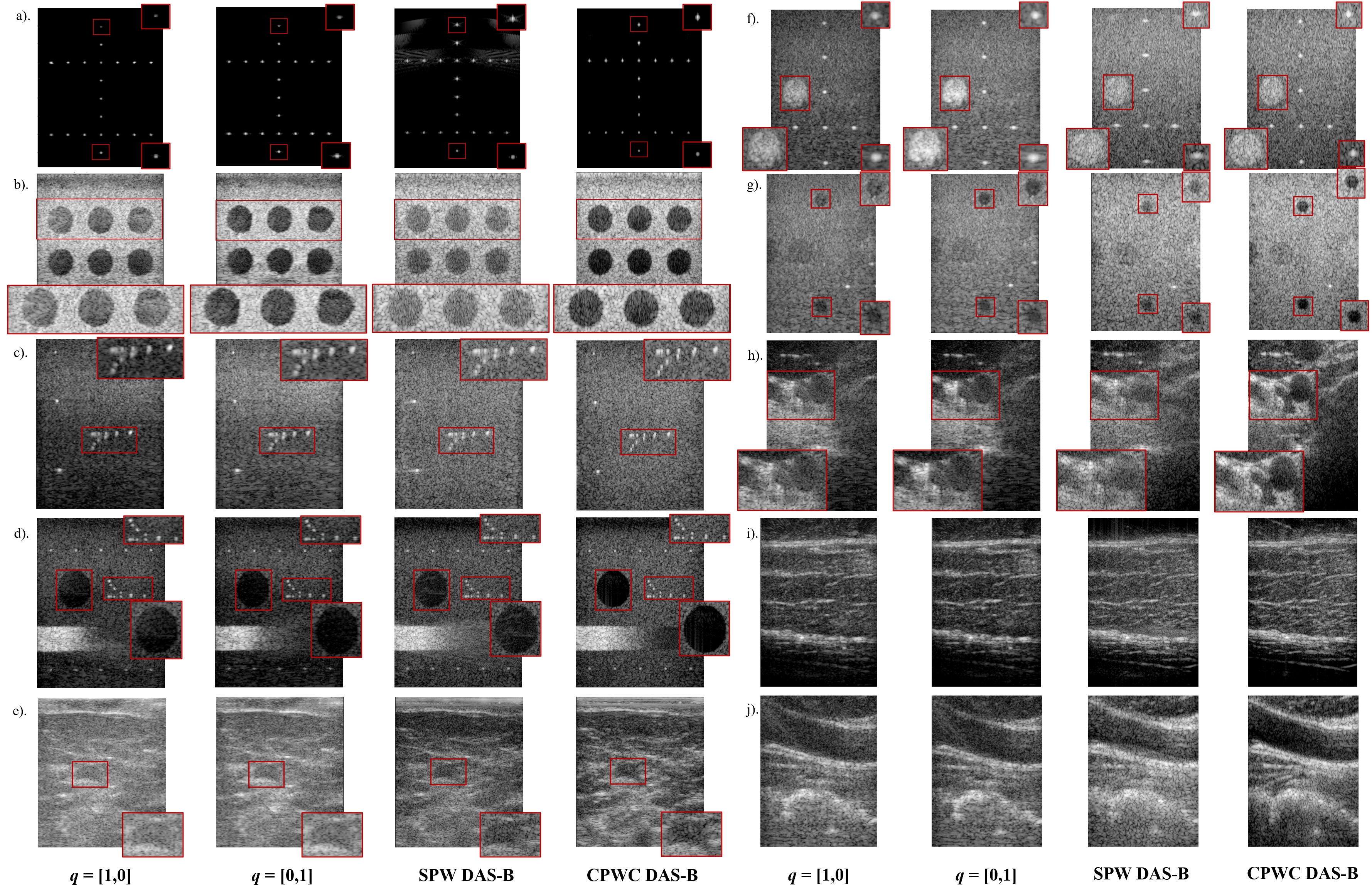}
    \caption{Results on PICMUS and CUBDL Datasets: a). In-silico Point Targets b). In-silico Targets c). In-vitro OSL002 d). In-vitro OSL010 e). In-vivo JHU024 f). In-vitro Point and Cyst Targets g). In-vitro Cyst Targets h). In-vivo Carotid Cross i). In-vivo Breast (JHU024) j). In-vivo Carotid Long}
    \label{fig:results}
\end{figure*}

\begin{table*}[!t]
\caption{Comparison with state-of-the-art beamforming methods on the PICMUS simulation dataset}
\label{tab:sota}
\centering
\footnotesize
\setlength{\tabcolsep}{4pt}
\renewcommand{\arraystretch}{1.1}
\begin{tabular}{lcccccccc}
\toprule
Method &
Input &
Lat. FWHM &
Ax. FWHM &
CNR &
gCNR &
SSIM &
PSNR \\
&
$\rightarrow$ Output &
(mm)$\downarrow$ &
(mm)$\downarrow$ &
(dB)$\uparrow$ &
$\uparrow$ &
$\uparrow$ &
(dB)$\uparrow$ \\
\midrule

DAS (SPW)
& 
& 0.793$\pm$0.123
& 0.413$\pm$0.007
& 5.76$\pm$0.39
& 0.839$\pm$0.026
& 0.456$\pm$0.462
& 16.18$\pm$5.78 \\

OATS~\cite{xiao2025optimized}
& ToF$\rightarrow$weights
& 0.699$\pm$0.202
& {0.401}$\pm${0.007}
& 4.54$\pm$0.80
& 0.711$\pm$0.045
& 0.287$\pm$0.411
& 13.88$\pm$8.10 \\

DATN~\cite{liu2026depth}
& ToF$\rightarrow$weights
& 0.538$\pm$0.122
& 0.410$\pm$0.007
& 6.41$\pm$0.80
& 0.882$\pm$0.097
& 0.405$\pm$0.451
& 15.37$\pm$8.02 \\

Hy-PCF~\cite{yu2026plane}
& ToF+Image$\rightarrow$weights
& 0.713$\pm$0.108
& 0.446$\pm$0.068
& 4.33$\pm$1.91
& 0.919$\pm$0.064
& 0.508$\pm$0.317
& 16.28$\pm$3.09 \\

\midrule

Proposed ($q=[1,0]$)
& ToF$\rightarrow$weights
& \textbf{0.506}$\pm$\textbf{0.111}
& \textbf{0.322}$\pm$\textbf{0.060}
& 5.95$\pm$0.87 
& 0.947$\pm$0.035
& \textbf{0.497}$\pm$\textbf{0.442}
& \textbf{18.42}$\pm$\textbf{6.13} \\

Proposed ($q=[0,1]$)
& ToF$\rightarrow$weights
& 0.519$\pm$0.136
& 0.358$\pm$0.075
& \textbf{6.47}$\pm$\textbf{1.38}
& \textbf{0.980}$\pm$\textbf{0.012}
& \textbf{0.703}$\pm$\textbf{0.324}
& \textbf{18.64}$\pm$\textbf{3.48} \\

\bottomrule
\end{tabular}
\end{table*}
Table~\ref{tab:picmus} gives the corresponding out-of-distribution results on the four annotated PICMUS phantoms for the same $\lambda_{\mathrm{metric}}{=}0.3$ model. Both imaging objectives outperform the SPW input in CNR and gCNR on both contrast phantoms. The contrast objective consistently provides higher contrast than the resolution objective, as expected. On the resolution phantoms, the ordering reverses as expected, with the resolution objective producing the narrower lateral FWHM on both sets. Both objectives also produce sharper axial responses than the $31$-angle CPWC target, while the SPW input retains an axial-resolution advantage because compounding broadens the axial response.

\begin{table*}[!ht]
\caption{Ablation study on the PICMUS simulation and experimental datasets.}
\label{tab:ablation}
\centering
\scriptsize
\setlength{\tabcolsep}{2pt}
\renewcommand{\arraystretch}{1.05}
\begin{tabular}{@{}l ccc | cc@{}}
\toprule
& \multicolumn{3}{c|}{\textbf{Simulation}}
& \multicolumn{2}{c@{}}{\textbf{Experimental}}\\
\cmidrule(lr){2-4}\cmidrule(l){5-6}
\textbf{Config.} & FWHM & gCNR & CNR & FWHM & gCNR \\
\midrule
\multicolumn{6}{@{}l}{\emph{Downsampling}}\\
Max-pool
 & $0.519{\pm}0.111$ & $0.860{\pm}0.104$ & $4.27{\pm}2.38$ & $0.879{\pm}0.174$ & $0.751{\pm}0.046$\\
Sym.\ S2D (2,2)
 & $0.512{\pm}0.123$ & $0.882{\pm}0.058$ & $3.91{\pm}2.37$ & $0.812{\pm}0.079$ & $0.774{\pm}0.065$\\
Aniso.\ S2D (4,2)
 & $\mathbf{0.512{\pm}0.117}$ & $\mathbf{0.888{\pm}0.084}$ & $\mathbf{4.46{\pm}2.74}$ & $\mathbf{0.687{\pm}0.074}$ & $\mathbf{0.785{\pm}0.004}$\\
\addlinespace
\multicolumn{6}{@{}l}{\emph{Kernel Shape}}\\
Square $3{\times}3$
 & $0.549{\pm}0.149$ & $0.888{\pm}0.084$ & $4.46{\pm}2.35$ & $0.879{\pm}0.174$ & $\mathbf{0.822{\pm}0.010}$\\
Axial $7{\times}3$
 & $0.544{\pm}0.121$ & $0.892{\pm}0.084$ & $4.48{\pm}2.56$ & $0.924{\pm}0.081$ & $0.789{\pm}0.024$\\
Lateral $3{\times}7$
 & $\mathbf{0.512{\pm}0.117}$ & $\mathbf{0.919{\pm}0.051}$ & $\mathbf{4.66{\pm}2.74}$ & $0.746{\pm}0.121$ & $0.785{\pm}0.004$\\
\addlinespace
\multicolumn{6}{@{}l}{\emph{Loss Functions}}\\
$\mathcal{L}_1$
 & $0.587{\pm}0.150$ & $0.852{\pm}0.087$ & $4.22{\pm}2.69$ & $1.012{\pm}0.173$ & $0.717{\pm}0.008$\\
SSIM
 & $0.721{\pm}0.306$ & $0.840{\pm}0.078$ & $\mathbf{5.95{\pm}1.10}$ & $1.220{\pm}0.118$ & $0.731{\pm}0.073$\\
Perceptual
 & $0.672{\pm}0.156$ & $0.853{\pm}0.095$ & $5.85{\pm}0.92$ & $1.134{\pm}0.154$ & $\mathbf{0.822{\pm}0.003}$\\
$\mathcal{L}_1$+Perc.
 & $0.528{\pm}0.124$ & $0.878{\pm}0.077$ & $3.62{\pm}2.22$ & $0.933{\pm}0.109$ & $0.609{\pm}0.124$\\
$\mathcal{L}_1$+Perc.+Met.
 & $\mathbf{0.512{\pm}0.117}$ & $\mathbf{0.888{\pm}0.084}$ & $4.46{\pm}2.74$ & $0.879{\pm}0.174$ & $0.785{\pm}0.004$\\
\addlinespace
\multicolumn{6}{@{}l}{\emph{Perceptual Backbone}}\\
SPW-CPWC clf.
 & $0.512{\pm}0.117$ & $0.888{\pm}0.084$ & $4.46{\pm}2.74$ & $0.770{\pm}0.148$ & $0.785{\pm}0.004$\\
VGG-13
 & $0.530{\pm}0.115$ & $0.884{\pm}0.087$ & $4.79{\pm}2.61$ & $0.879{\pm}0.174$ & $0.776{\pm}0.027$\\
ResNet-50
 & $\mathbf{0.506{\pm}0.111}$ & $\mathbf{0.916{\pm}0.067}$ & $\mathbf{4.75{\pm}2.87}$ & $0.845{\pm}0.096$ & $\mathbf{0.791{\pm}0.011}$\\
\addlinespace
\multicolumn{6}{@{}l}{\emph{Conditioning Placement}}\\
Decoder only
 & $\mathbf{0.512{\pm}0.117}$ & $\mathbf{0.888{\pm}0.084}$ & $\mathbf{4.46{\pm}2.74}$ & $0.825{\pm}0.205$ & $0.785{\pm}0.004$\\
Encoder+Decoder
 & $0.528{\pm}0.119$ & $0.875{\pm}0.098$ & $4.46{\pm}2.34$ & $0.879{\pm}0.174$ & $0.746{\pm}0.130$\\
Bottleneck+Decoder
 & $0.520{\pm}0.149$ & $0.861{\pm}0.107$ & $4.16{\pm}2.50$ & $0.841{\pm}0.167$ & $0.775{\pm}0.042$\\
\bottomrule
\end{tabular}
\end{table*}

\begin{figure}
    \centering
    \includegraphics[width=\linewidth]{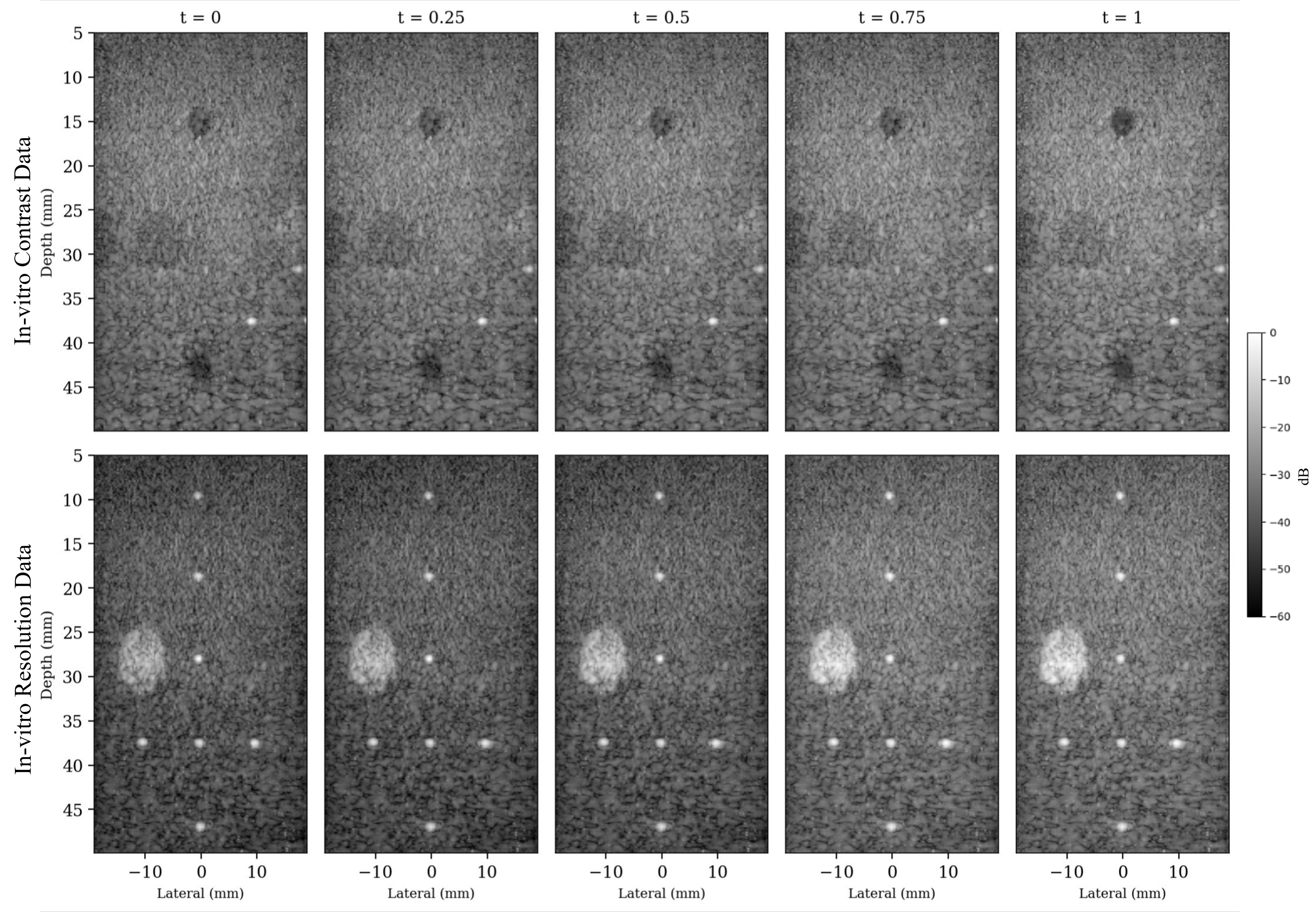}
    \caption{Continuous control via objective interpolation. Sweeping the
conditioning vector $q(t)=[1-t,\,t]$ from the resolution endpoint ($t{=}0$)
to the contrast endpoint ($t{=}1$) produces a smooth, monotonic transition
in lateral resolution and lesion contrast, with intermediate $t$ giving
operating points between the two supervised objectives. This indicates a smooth interpolation between the two supervised operating points rather than a switch between two discrete modes.}
    \label{fig:intent_continuum}
\end{figure}

Table~\ref{tab:ablation} reports an ablation on the PICMUS simulation and
experimental sets. The anisotropic S2D$(4,2)$ downsampling and decoder-only
conditioning give the best experimental FWHM and gCNR, and the metric loss
is decisive for objective-specific behaviour: adding it to the
$\mathcal{L}_1$+$\mathcal{L}_{perc}$ loss in the model is what activates the intended
resolution-contrast control. The lateral $3\times7$ kernel provides the strongest overall simulation
performance and improves experimental FWHM relative to the square kernel,
supporting its use for the final architecture.

\section{Discussion}
\label{sec:discussion}
The proposed framework formulates the resolution-contrast trade-off in ultrasound receive beamforming as a controllable learning problem. Conditioned on a physically defined receive-aperture regime, the proposed network maps the same single-plane-wave channel data to objective-dependent apodization weights while retaining conventional delay-and-sum beamforming. Changing the conditioning state produces distinct behaviours: the resolution objective narrows lateral point-target responses, whereas the contrast objective improves lesion-background separation. Interpolating the conditioning variable yields a smooth family of intermediate
reconstructions (Fig.~\ref{fig:intent_continuum}), with $t\in[0,1]$ moving
between the two supervised endpoints. Since intermediate values are not
independently supervised, this behavior is interpreted as interpolation
between operating regimes rather than independent validation of intermediate
objectives.

The conditioning variable is physically anchored to receive-aperture configurations through the receive ($F\#$), giving the supervised operating points and their interpolation a clear interpretation. This constraint supports cross-domain generalization: although trained only on simulations, the model preserves its objective-dependent behaviour on independent PICMUS and CUBDL acquisitions with different probes and acquisition conditions. While it does not match the compounded reference on every metric, its consistent behaviour on unseen data suggests that learning apodization provides a useful physical constraint for transfer across domains.

The formulation is practical: a 24.6M-parameter single network reconstructs full frames at ~10.8 fps on an NVIDIA RTX 4070, with conditioning replacing separate beamforming networks at no added parameter cost.

The study has three main limitations. First, the conditioning space covers only resolution- and contrast-oriented objectives. Although interpolation yields meaningful intermediate behavior, it does not span the full range of clinically relevant imaging tasks. Second, the physics-guided losses are differentiable surrogates that capture only selected aspects of image quality, and additional statistical or task-specific criteria may better characterize tissue and lesion structure. Third, validation is limited to PICMUS and CUBDL data and a fixed linear probe formulation. Evaluation across more probes, systems, anatomies, and clinical tasks is therefore needed before clinical deployment.

The framework could be extended by conditioning on downstream tasks such as segmentation or detection, by replacing the predefined vector with a task representation derived from text or speech, and by adding differentiable losses targeting texture, boundary preservation, or clutter suppression. Ultimately, a shared task-conditioned beamformer could map a task description to an apodization strategy optimized by downstream performance, making beamforming an adaptive, task-aware component of the ultrasound pipeline.
\section{Conclusion}
\label{sec:conclusion}
We presented a physically grounded, objective-conditioned deep learning-based ultrasound beamformer that predicts receive-apodization functions from non-steered plane-wave channel data. The proposed framework enables continuous control of the resolution-contrast trade-off without separate models or retraining. Results on unseen PICMUS and CUBDL data show that this behavior transfers beyond the simulated training distribution through interpretable aperture changes rather than arbitrary image transformations. The framework provides a basis for a shared task-conditioned beamformer that could incorporate downstream objectives, richer differentiable criteria, and structured or natural-language conditioning.

\section*{Acknowledgement}
The authors would like to thank Dr. Timothy Liu and the NVIDIA AI Technology Center (NVAITC) Singapore for their valuable support and guidance throughout the course of this work. Generative AI tools were used for brainstorming and language editing. All technical claims, experiments, and code were authored and verified by the authors.

\bibliographystyle{IEEEtran}
\bibliography{references}

@article{montaldo2009coherent,
  title={Coherent plane-wave compounding for very high frame rate ultrasonography and transient elastography},
  author={Montaldo, Gabriel and Tanter, Micka{\"e}l and Bercoff, J{\'e}r{\'e}my and Benech, Nicolas and Fink, Mathias},
  journal={IEEE transactions on ultrasonics, ferroelectrics, and frequency control},
  volume={56},
  number={3},
  pages={489--506},
  year={2009},
  publisher={IEEE}
}

@article{tanter2014ultrafast,
  title={Ultrafast imaging in biomedical ultrasound},
  author={Tanter, Mickael and Fink, Mathias},
  journal={IEEE transactions on ultrasonics, ferroelectrics, and frequency control},
  volume={61},
  number={1},
  pages={102--119},
  year={2014},
  publisher={IEEE}
}

@article{sandrin1999time,
  title={Time-resolved pulsed elastography with ultrafast ultrasonic imaging},
  author={Sandrin, Laurent and Catheline, Stefan and Tanter, Micka{\"e}l and Hennequin, Xavier and Fink, Mathias},
  journal={Ultrasonic imaging},
  volume={21},
  number={4},
  pages={259--272},
  year={1999},
  publisher={SAGE Publications Sage CA: Los Angeles, CA}
}

@article{luchies2018deep,
  title={Deep neural networks for ultrasound beamforming},
  author={Luchies, Adam C and Byram, Brett C},
  journal={IEEE transactions on medical imaging},
  volume={37},
  number={9},
  pages={2010--2021},
  year={2018},
  publisher={IEEE}
}

@article{couture2012ultrasound,
  title={Ultrasound contrast plane wave imaging},
  author={Couture, Olivier and Fink, Mathias and Tanter, Mickael},
  journal={IEEE transactions on ultrasonics, ferroelectrics, and frequency control},
  volume={59},
  number={12},
  pages={2676--2683},
  year={2012},
  publisher={IEEE}
}

@article{perrot2021so,
  title={So you think you can DAS? A viewpoint on delay-and-sum beamforming},
  author={Perrot, Vincent and Polichetti, Maxime and Varray, Fran{\c{c}}ois and Garcia, Damien},
  journal={Ultrasonics},
  volume={111},
  pages={106309},
  year={2021},
  publisher={Elsevier}
}

@inproceedings{trahey1986speckle,
  title={Speckle SNR vs. lateral resolution in ultrasound imaging: What's the tradeoff?},
  author={Trahey, GE and Smith, SW and Carroll, BA and von Ramm, OT},
  booktitle={IEEE 1986 Ultrasonics Symposium},
  pages={811--814},
  year={1986},
  organization={IEEE}
}

@article{gasse2017high,
  title={High-quality plane wave compounding using convolutional neural networks},
  author={Gasse, Maxime and Millioz, Fabien and Roux, Emmanuel and Garcia, Damien and Liebgott, Herv{\'e} and Friboulet, Denis},
  journal={IEEE transactions on ultrasonics, ferroelectrics, and frequency control},
  volume={64},
  number={10},
  pages={1637--1639},
  year={2017},
  publisher={IEEE}
}

@article{luijten2020adaptive,
  title={Adaptive ultrasound beamforming using deep learning},
  author={Luijten, Ben and Cohen, Regev and De Bruijn, Frederik J and Schmeitz, Harold AW and Mischi, Massimo and Eldar, Yonina C and Van Sloun, Ruud JG},
  journal={IEEE Transactions on Medical Imaging},
  volume={39},
  number={12},
  pages={3967--3978},
  year={2020},
  publisher={IEEE}
}

@article{hyun2019beamforming,
  title={Beamforming and speckle reduction using neural networks},
  author={Hyun, Dongwoon and Brickson, Leandra L and Looby, Kevin T and Dahl, Jeremy J},
  journal={IEEE transactions on ultrasonics, ferroelectrics, and frequency control},
  volume={66},
  number={5},
  pages={898--910},
  year={2019},
  publisher={IEEE}
}

@article{khan2021switchable,
  title={Switchable and tunable deep beamformer using adaptive instance normalization for medical ultrasound},
  author={Khan, Shujaat and Huh, Jaeyoung and Ye, Jong Chul},
  journal={IEEE Transactions on Medical Imaging},
  volume={41},
  number={2},
  pages={266--278},
  year={2021},
  publisher={IEEE}
}

@article{li2026dp,
  title={DP-HAFNet: A dual-path hierarchical adaptive fusion network for ultrasound image reconstruction from single plane-wave RF data},
  author={Li, Jiajin and Sun, Wenwen and Wang, Yuanguo and Zheng, Chichao and Han, Zhihui and Peng, Hu and Zhang, Chaoxue and Zhou, Jinhua and Wang, Yadan},
  journal={Measurement},
  pages={121562},
  year={2026},
  publisher={Elsevier}
}

@article{amar2025deep,
  title={Deep Task-Based Beamforming and Channel Data Augmentations for Enhanced Ultrasound Imaging},
  author={Amar, Ariel and Grubstein, Ahuva and Atar, Eli and Peri-Hanania, Keren and Glazer, Nimrod and Rosen, Ronnie and Savariego, Shlomi and Eldar, Yonina C},
  journal={arXiv preprint arXiv:2502.00524},
  year={2025}
}

@article{yu2026plane,
  title={Plane-wave image reconstruction with Hy-PCF: A Novel Hybrid CNN-Transformer for Progressive Cross-domain Fusion},
  author={Yu, Yanyan and Wang, Tianli and Qiang, Yu and Wang, Xingying and Chen, Xin and Qiu, Weibao},
  journal={IEEE Transactions on Medical Imaging},
  year={2026},
  publisher={IEEE}
}

@inproceedings{perez2018film,
  author    = {Ethan Perez and Florian Strub and Harm de Vries and Vincent Dumoulin and Aaron Courville},
  title     = {FiLM: Visual Reasoning with a General Conditioning Layer},
  booktitle = {Proceedings of the AAAI Conference on Artificial Intelligence},
  volume    = {32},
  number    = {1},
  year      = {2018}
}

@article{ho2022classifier,
  title={Classifier-free diffusion guidance},
  author={Ho, Jonathan and Salimans, Tim},
  journal={arXiv preprint arXiv:2207.12598},
  year={2022}
}

@inproceedings{garcia2021make,
  title={Make the most of MUST, an open-source Matlab UltraSound Toolbox},
  author={Garcia, Damien},
  booktitle={2021 IEEE international ultrasonics symposium (IUS)},
  pages={1--4},
  year={2021},
  organization={IEEE}
}

@inproceedings{liebgott2016plane,
  title={Plane-wave imaging challenge in medical ultrasound},
  author={Liebgott, Herve and Rodriguez-Molares, Alfonso and Cervenansky, Fr{\'e}deric and Jensen, J{\o}rgen Arendt and Bernard, Olivier},
  booktitle={2016 IEEE International ultrasonics symposium (IUS)},
  pages={1--4},
  year={2016},
  organization={IEEE}
}

@inproceedings{bell2020challenge,
  title={Challenge on ultrasound beamforming with deep learning (CUBDL)},
  author={Bell, Muyinatu A Lediju and Huang, Jiaqi and Hyun, Dongwoon and Eldar, Yonina C and Van Sloun, Ruud and Mischi, Massimo},
  booktitle={2020 IEEE International Ultrasonics Symposium (IUS)},
  pages={1--5},
  year={2020},
  organization={IEEE}
}

@inproceedings{shi2016real,
  title={Real-time single image and video super-resolution using an efficient sub-pixel convolutional neural network},
  author={Shi, Wenzhe and Caballero, Jose and Husz{\'a}r, Ferenc and Totz, Johannes and Aitken, Andrew P and Bishop, Rob and Rueckert, Daniel and Wang, Zehan},
  booktitle={Proceedings of the IEEE conference on computer vision and pattern recognition},
  pages={1874--1883},
  year={2016}
}

@inproceedings{shang2016understanding,
  title={Understanding and improving convolutional neural networks via concatenated rectified linear units},
  author={Shang, Wenling and Sohn, Kihyuk and Almeida, Diogo and Lee, Honglak},
  booktitle={international conference on machine learning},
  pages={2217--2225},
  year={2016},
  organization={PMLR}
}

@inproceedings{johnson2016perceptual,
  title={Perceptual losses for real-time style transfer and super-resolution},
  author={Johnson, Justin and Alahi, Alexandre and Fei-Fei, Li},
  booktitle={European conference on computer vision},
  pages={694--711},
  year={2016},
  organization={Springer}
}

@article{xiao2025optimized,
  title={Optimized Apodizations with Training Simulations (OATS): Learned depth-dependent apodizations via differentiable beamforming for reduced operator tuning},
  author={Xiao, Di and Nahas, Hassan and Hiroshima, Misaki and Kishimoto, Aya and Yu, Alfred CH},
  journal={Ultrasonics},
  pages={107827},
  year={2025},
  publisher={Elsevier}
}

@article{liu2026depth,
  title={Depth-aware transformer network for adaptive beamforming in single-angle plane-wave imaging},
  author={Liu, Peng and Zhao, Wanchen and Wang, Yachen and Wei, Yihang and Fan, Shangchun and Xing, Weiwei and Qu, Xiaolei},
  journal={Measurement},
  pages={121721},
  year={2026},
  publisher={Elsevier}
}

@inproceedings{savoia2014improved,
  title={Improved lateral resolution and contrast in ultrasound imaging using a sidelobe masking technique},
  author={Savoia, Alessandro Stuart and Caliano, Giosu{\`e} and Matrone, Giulia and Magenes, Giovanni and Ramalli, Alessandro and Boni, Enrico and Tortoli, Piero},
  booktitle={2014 IEEE International Ultrasonics Symposium},
  pages={1682--1685},
  year={2014},
  organization={IEEE}
}

@article{holmes2024f,
  title={F-number optimization for synthetic aperture delay-multiply-and-sum reconstruction},
  author={Holmes, Philip M and Lee, Hyoung-Ki and Urban, Matthew W},
  journal={Ultrasonics},
  volume={136},
  pages={107158},
  year={2024},
  publisher={Elsevier}
}

@inproceedings{gopikrishnan2026adapt,
  author    = {Gopika Gopikrishnan and Mahesh Raveendranatha Panicker and
               Timothy Liu and Ng Aik Beng and Simon Chong-Wee See},
  title     = {ADAPT: Multibeamformer with Tunable Weight Fusion and Patch-Wise Learning for Ultrasound Imaging},
  booktitle = {2026 IEEE International Symposium on Biomedical Imaging (ISBI)},
  year      = {2026},
  address   = {London, United Kingdom}
}

\end{document}